\documentclass[epj] {svjour} 

\usepackage{amsmath}
\usepackage{amssymb}
\usepackage[dvips] {rotating}
\usepackage{epsf}
\usepackage{psfig}
\usepackage{epsfig}
\usepackage[mathscr]{eucal}
\newcommand{\bm}[1]{\mbox{\boldmath$#1$}}

\newlength{\colwidth}
\newcommand{\beqs}{\begin{eqnarray*}}
\newcommand{\eeqs}{\end{eqnarray*}}
\newcommand{\beq}{\begin{eqnarray}}
\newcommand{\eeq}{\end{eqnarray}}

\newcommand{\bc}{\begin{center}}
\newcommand{\ec}{\end{center}}
\newcommand{\cd}{\cdot}
\newcommand{\p}{\partial}
\newcommand{\bt}{\mathbf{t}}
\newcommand{\bu}{\mathbf{u}}
\newcommand{\br}{\mathbf{r}}
\newcommand{\Ds}{D_{\perp}}
\newcommand{\Dp}{D_{\parallel}}
\newcommand{\n}{\nabla}
\newcommand{\al}{\alpha}

\begin {document}
%
%
%
\title{Nonlinear competition between asters and stripes
 in filament--motor--systems}

%
%
\author{Falko Ziebert   and   Walter Zimmermann         }
\titlerunning{Nonlinear competition between asters and stripes}
\authorrunning{F.~Ziebert and W.~Zimmermann}

%
%
\institute{Theoretische Physik, Universit\"at des Saarlandes,
           D-66041 Saarbr\"ucken, Germany
}

%
%
\abstract{
A model for polar filaments interacting via molecular motor
complexes is investigated which exhibits bifurcations to spatial patterns.
It is shown
that the homogeneous distribution of filaments, such as actin or
microtubules,  may become either unstable
with respect to an orientational instability
of a finite wave number
or with respect to modulations of the filament density, where
long wavelength modes are amplified as well.
Above threshold nonlinear interactions
select either stripe patterns or periodic asters. The existence and stability
ranges of each  pattern close to threshold are
predicted in terms of a weakly nonlinear perturbation analysis,
which is confirmed by numerical simulations of the
basic model equations.
The two relevant parameters determining the bifurcation scenario of
the model can be related to the concentrations of the active molecular motors
and of the filaments respectively,
which both could be easily regulated by the cell.
}
\PACS { {87.16.-b}{Subcellular structure} \and
        {47.54.+r}{Pattern formation} \and
        {89.75.-k}{Complex systems}
}
%
%
%
\maketitle
%

%
%
%

\section{Introduction}
\label{Intro}
In eukaryotic cells the polar filaments actin  and microtubules
interacting with motor proteins play a crucial
role in intracellular organisation and transport as well as for the static and
dynamical  structure of the cytoskeleton \cite{Alberts:2001,Howard:2001}.
Most prominently, microtubules and kinesin are involved in highly connected
dynamical structures, such as the mitotic
spindle in cell division \cite{Karsenti:86.1,Karsenti:01.1}, while the motility of the cell as a whole
is governed by acto-myosin complexes \cite{Pantaloni:01.1,Borisy:98.1}.
The dynamical behavior of such assemblies depends
on the available
biological fuel  ATP (Adenosine-triphosphate),
which is consumed during the motion of motor proteins along
the filaments. Vesicles  for instance are
transported across a cell by motors moving
 along the  tracks defined
by microtubules, or -- which is the scope of this work -- oligomeric
motor proteins that attach to two or more filaments
induce relative motion between
 neighboring filaments and cause dynamical networks.
The latter process is vitally important in cells, since the
cytoskeleton constituted of the filaments has to be self-organized
and even actively reorganized not only during
cell--locomotion but also in order to react to outer stimuli (chemotaxis).
During mitosis, microtubules attach to the chromosomes,
which are then divided
and the two halves finally are transported by
the motor--induced filament sliding into the two evolving daughter cells.

Since the situation is very complex
in a living cell,
well designed in vitro experiments
are the agent of choice for controlled explorations of
prominent aspects  of cellular systems. Recent
experimental progress yielded indeed important insights
into organization and
dynamical properties of the cell, which in turn call
for modeling ac\-ti\-vi\-ties to foster their  deeper understanding.
These experiments comprise investigations of self-organization in
fi\-la\-ment--mo\-tor mixtures  of microtubules in the presence of
a single type of motor protein \cite{Urrutia:1991.1,Nedelec:1997.1,Surrey:98.1}
and more recently  also in actin--myosin networks \cite{Kaes:2002.1,Smith:04.1}
as well as assays where two types of motors interact with microtubules  \cite{Leibler:2001.1}.
Even in such model systems, simple compared to a living cell, there has been found a great variety
of different two-dimensional patterns, such as
stripe patterns, asters, vortices and irregular arrangements.

Works  on modeling  active fi\-la\-ment--mo\-tor sys\-tems
comprise molecular dynamics simulations
\cite{Nedelec:1997.1,Leibler:2001.1,Nedelec:2002.1} and
 mean field appro\-xi\-ma\-tions for spin-like models displaying
stripe-like patterns \cite{Bassetti:2000.1}.
More recently a phenomenological model of a motor density interacting with a phenomenological vector field
has been considered which is able to reproduce asters and vortex-like solutions
\cite{Kardar:2001.1} as well as  models for
bundle contraction in one spatial dimension  \cite{Sekimoto:1996.1,Kruse:2000.1,Kruse:2003.1}.

Here we follow a more microscopic approach starting
from the Smoluchowski  equation for the  spatial and
angular distribution of rigid rods \cite{DoiEd}, which approximate
the rather stiff microtubules and with limitations also the actin filaments.
It is  well known since  Onsager's theory that
an ideal rigid rod system exhibits a nematic liquid crystalline order
beyond some critical rod density
\cite{Onsager:49.1,deGennes:93} which is still present in generalizations for
semiflexible polymers, albeit at a slightly higher filament density \cite{Chen:93.1},
and has been observed in vitro for both kinds of biopolymers
\cite{Hitt:90.1,Suzuki:91.1}.
To use Onsager's equilibrium argument of excluded volume induced transitions in a nonequilibrium
system like the cell, one can formulate phenomenological models as done recently
to describe how
nucleation and decay of the filaments in a cell at a density
close to the isotropic--nematic transition  may give rise to
spatially periodic patterns \cite{Ziebert:2004.1}.
Alternatively one can use the aforementioned Smoluchowski equation for rigid rods
\cite{DoiEd}, which is an approach widely used in polymer and colloidal science
and recently has been supplemented by active
currents to describe filament--motor--systems \cite{Liverpool:2003.2} in a way inspired
by a macroscopic model for filament bundling in
one spatial  dimension \cite{Kruse:2000.1}.
In this approach one makes use of the fact that the small motor proteins
diffuse much faster than the filaments.
Hence the density of the motors can be assumed to be homogeneous and -- as well as
properties like the mean velocity and the duty ratio of the motors --
enters into the model only via the coefficients. The phenomenological description
of the active currents can be derived by
symmetry considerations and includes three major contributions,
whereof in this work we focus on the effects on pattern
formation of only two of them.
Since most of the  experimental assays are quasi two-dimensional, we restrict
our analysis also to two spatial dimensions.

The filament-filament interactions induced by the motors as well
as by excluded volume effects are nonlocal. Thus they have to be approximated
by a gradient expansion, which must not be
truncated at the leading order, because on their basis
the instabilities of the homogeneous filament
distributions cannot be predicted as pointed out recently \cite{Ziebert:2004.2},
but has to be continued up to
fourth order derivatives.
Since in vitro aster-like patterns evolve at much lower filament density than the
isotropic--nematic transition, a moment expansion of the probability density function
can be truncated to derive a closed set of
equations for the physical observables,
which are the density and
the orientation field of the filaments.
After a detailed linear analysis the major goal of this work is
a characterization of  the nonlinear behavior of
patterns beyond  their threshold.
For this purpose we employ on the one hand numerical simulations
of the coupled equations for the filament density
and the orientational field.
On the other hand we use the  powerful method of
amplitude expansion, where equations of motion for
the amplitudes of the spatially periodic pattern  are
derived close to the pattern forming instability \cite{crossho:93,Newell:92.1,Manneville:90}.

The generic patterns  in  two--dimensional
systems close to threshold are
stripes, squares or hexagonal patterns
and for  the amplitudes of these patterns generic nonlinear amplitude equations
may be derived \cite{crossho:93,Ciliberto:90.1}.
In numerical simulations
we found either stripes or squares (the absence of hexagonal structures
can be explained by the hierarchy of equations in the derivation of the amplitude
equation)
and therefore
we focus on the equations for
stripe and square patterns (as  described by Eq.~(\ref{ansatz}) below),
which are of the  form \cite{crossho:93,Segel:65.1}
\begin{subequations}
\label{Aeq_int}
\beq \label{Aeq1_int}
\tau_0\p_t X &=&\varepsilon X - g_1|X|^2 X - g_2 |Y|^2 X\enspace,\\
\label{Aeq2_int}
 \tau_0\p_t Y &=&\varepsilon Y - g_1|Y|^2 Y - g_2 |X|^2 Y\enspace.
\eeq
\end{subequations}
Here $\varepsilon$ denotes the deviation of the control parameter
from its value at  the threshold
of pattern formation, while $X$ and $Y$ describe the amplitudes of a spatially periodic pattern in one (stripe) or
two orthogonal directions (square), namely the  $x$- and $y$-direction.
These amplitude equations are universal for all pattern forming systems of a given symmetry
and the specific information of each
system is only encoded in the coefficients, namely $\tau_0, g_1, g_2$,
which are functions of the parameters of the underlying system under consideration.
The nonlinear analysis of  Eqs.~(\ref{Aeq_int})
is worked out  analytically and by this
advantage a  whole phase diagram for
the various  patterns close to threshold is presented.
In contrast, such  an investigation
would be very cumbersome and time consumptive
using numerical simulations of the model equations.
The solution $|X|= |Y|$ corresponds to a square pattern for the orientational field of the filaments
which resembles very much a structure of periodically arranged asters.

The work is organized as follows.
The underlying microscopic model for the filament distribution
 is presented in Sec.~\ref{model} and in the same section
 we derive also the coupled
equations for the density and orientation of the filaments,
whereby the technical parts are moved to
the appendices. The thresholds of the pattern forming
instabilities and its dependence of the parameters
are determined in section \ref{thresh}, where we find
an instability leading to density modulations,
an orientational instability as well as
a liquid-crystalline isotropic--nematic (I--N) transition.
With regard to the formation of asters, the orientational instability
is the most interesting one and its nonlinear  behavior, namely
the nonlinear stripe solutions and the periodic lattice of asters and inverse asters as well as
 their stability
are investigated in more detail in
 Secs. \ref{weaknonlin} and \ref{numsim}.
The work is finished
with a summary and concluding remarks.

\section{The model}
\label{model}
The probability of finding a rod (with fixed length $L$) at the position
${\bf r}$  with the orientation
$\bf{u}$ (with $|\bu|=1$) at time  $t$  is described by the
distribution function $\Psi(\br,\bu,t)$\footnote{In the following we will write $\Psi(\br,\bu)$ for
reasons of brevity and
$\Psi(\br,\bu,t)$ only if we want to emphasize the time dependence.}
and the governing so-called Smoluchowski equation is just the continuity equation
for the probability
\beq\label{conserved}
\partial_t\Psi+\nabla\cdot J_t+\mathcal{R}\cdot J_r=0 \enspace.
\eeq
Homogeneously distributed molecular motors
are supposed to interact with the rods
inducing active currents on them.
The total translational and rotational
currents are given in Cartesian coordinates by
\begin{subequations}
\label{current}
\beq\label{tcurrent}
J_{t,i}&=&-D_{ij}\left[\partial_j\Psi+\Psi\partial_j V_{ex}\right] +J_{t,i}^{a}\enspace,\\
\label{rcurrent}
J_{r,i}&=&-D_r\left[\mathcal{R}_i\Psi+\Psi\mathcal{R}_i V_{ex}\right] +J_{r,i}^{a}\enspace.
\eeq
\end{subequations}
The terms with an upper index $a$ are the motor--mediated
active currents,  the translational diffusion matrix reads
\beq
D_{ij}=D_{\parallel}u_{i}u_{j}+D_{\perp}\left(\delta_{ij}-u_{i}u_{j}\right),
\eeq
$D_r$ is the rotational diffusion constant
and the operator of rotational diffusion is given by
\beq\label{rotop}
\mathcal{R}=\mathbf{u}\times\partial_{\mathbf{u}}\enspace.
\eeq
$V_{ex}$ describes the excluded volume interaction
\beq\label{Vex}
V_{ex}(\mathbf{r},\mathbf{u})=\int d\mathbf{u'}\int d\mathbf{r'} ~W(\mathbf{r\hspace{-1mm}-\hspace{-1mm}r'},\mathbf{u},\mathbf{u'})\Psi(\mathbf{r'},\mathbf{u'})
\eeq
between rods, where
the interaction kernel is expressed in terms of the so-called
Straley coordinates   $\zeta$
and $\eta$  in two dimensions \cite{Straley:1973.2}
\beq\label{kernel}
W(\mathbf{r\hspace{-1mm}-\hspace{-1mm}r'},\mathbf{u},\mathbf{u'})=|\bu\hspace{-1mm}\times\hspace{-1mm}\bu'|\hspace{-2mm}\int\limits_{-L/2}^{L/2}\hspace{-2mm}d\zeta\hspace{-2mm}\int\limits_{-L/2}^{L/2}\hspace{-2mm}
d\eta\hspace{1mm}\delta\left(\br\hspace{-1mm}-\hspace{-1mm}\br'\hspace{-1mm}+\hspace{-1mm}\bu\zeta\hspace{-1mm}+\hspace{-1mm}\bu'\eta\right).
\eeq
This expression takes into account
that there is only an interaction between rods
with  coordinates
 $(\br,\bu)$ and
$(\br',\bu')$ in the case of a finite overlap
 i.e. if the connection vector $\br-\br'$
can be constructed by a linear combination $\bu\zeta+\bu'\eta$
of the rod orientations
with $-L/2 < \zeta,\eta < L/2$.
The active currents induced  by the  homogeneous motor density
are
\beq\label{acurrent}
& &\hspace{-6mm}J^{a}_t\hspace{-1mm}=\Psi(\mathbf{r},\mathbf{u})\hspace{-2mm}\int\hspace{-1mm}d\mathbf{u'}\hspace{-2mm}\int\hspace{-1mm}
d\mathbf{r'}\mathbf{v}(\mathbf{r\hspace{-1mm}-\hspace{-1mm}r'}\hspace{-1mm},\mathbf{u},\mathbf{u'}) W(\mathbf{r\hspace{-1mm}-\hspace{-1mm}r'}\hspace{-1mm},\mathbf{u},\mathbf{u'}) \Psi(\mathbf{r'}\hspace{-1mm},\mathbf{u'})\,,\nonumber \\
\label{arcurrent}
& &\hspace{-6mm}J^{a}_r\hspace{-1mm}=\Psi(\mathbf{r},\mathbf{u})\hspace{-2mm}\int\hspace{-1mm}d\mathbf{u'}\hspace{-2mm}\int\hspace{-1mm}
d\mathbf{r'}\bm{\omega}(\mathbf{u},\mathbf{u'}) W(\mathbf{r\hspace{-1mm}-\hspace{-1mm}r'}\hspace{-1mm},\mathbf{u},\mathbf{u'}) \Psi(\mathbf{r'}\hspace{-1mm},\mathbf{u'})
\eeq
with the translational and rotational velocity
given by
\beq\label{vel}
\mathbf{v}(\mathbf{r\hspace{-1mm}-\hspace{-1mm}r'}\hspace{-1mm},
\mathbf{u},\mathbf{u'})&=&\frac{\alpha}{2}\frac{\mathbf{r-r'}}{L}
\frac{1+\mathbf{u}\cdot\mathbf{u'}}{|\mathbf{u}\times\mathbf{u'}|}+
\frac{\beta}{2}\frac{\mathbf{u'-u}}{|\mathbf{u}\times\mathbf{u'}|}\enspace,\\
\label{omegel}
\bm{\omega}(\mathbf{u},\mathbf{u'})&=&\gamma(\mathbf{u}\cdot\mathbf{u'})\frac{\mathbf{u}\times\mathbf{u'}}{|\mathbf{u}\times\mathbf{u'}|}\enspace,
\eeq
respectively.
The active currents are normalized to the interaction volume
and they fulfill both,
the conservation of translational and rotational
momentum in the absence of
external forces and torques, as well as translational and rotational
invariance (c.f. \cite{Liverpool:2003.2} and appendix~\ref{symmvel}).

According to its functional dependence  $1 +\mathbf{u}\cdot\mathbf{u'}$,
the first term  in Eq.~(\ref{vel}) contributes at maximum
to the filament transport for a parallel orientation, which identifies it as the leading
pattern forming mechanism in counteraction with the diffusive motion.
It arises from a difference in motor activity between the filament center and the filament endpoints,
where motors stall for some finite time $\tau_f$, which has been recognised as a crucial
condition for aster formation \cite{Nedelec:2001.1}. It depends also on the relative filament
separation $\br-\br'$,
hence no relative motion is obtained if the motor connects two filaments having the same centers of mass.

The second term    in Eq.~(\ref{vel}),  proportional
to $\beta$, causes  filament translations along $\bu'-\bu$ and
therefore becomes maximal for antiparallel filament orientations.
The contribution  in Eq.~(\ref{omegel}) causes rotational currents.
In this work we focus mainly on active currents
caused by the $\alpha$-term and especially on its
effects on pattern formation in two spatial dimensions.

Eq.~(\ref{conserved}) together with the currents as defined by
Eqs.~(\ref{current}) and (\ref{acurrent})
cannot be solved analytically for $\Psi(\mathbf{r},\mathbf{u},t)$
and even numerically, being a non\-linear
in\-te\-gro-differen\-tial e\-qua\-tion, it would be rather  intricate.
However, these equations may be simplified by
expressing the distribution function in terms of
its moments.
For a rod distribution, the zeroth, first and second moment with respect to the
orientation
correspond to the filament density $\rho({\bf r},t)$, the polar orientation $\bt ({\bf r},t)$
and the nematic order parameter $S_{ij}({\bf r},t)$,
which are defined by the following expressions
\beq
\label{moments}
\rho(\br,t)&=&\int d\bu\enspace\Psi(\br,\bu,t)\enspace,\nonumber\\
\bt(\br,t)&=&\int d\bu~\bu~ \Psi(\br,\bu,t)\enspace,\nonumber\\
S_{ij}(\br,t)&=&\int d\bu ~u_i u_j~\Psi(\br,\bu,t)\enspace.
\eeq
In contrast to a usual lyotropic liquid crystal \cite{Onsager:49.1},
here the filaments are polar with respect to the
motor action which breaks the $\pm\bu$-symmetry and therefore
the first moment may become finite.

The critical filament density of the homogeneous iso\-tro\-pic--ne\-ma\-tic (I--N)
transition can be determined by the
stability of the second moment of the rotational diffusion contribution, i.e. where
$-D_r\int u_\alpha u_\beta\mathcal{R}\left[\mathcal{R}\Psi+\Psi\mathcal{R} V_{ex}\right]$
changes its sign.
This condition
yields the critical density
\beq\label{rhoIN}
\rho_{IN}=\frac{3}{2}\pi
\eeq
in two spatial dimensions where
the I--N transition is of second order.

In order to obtain a closed set of
equations of motion, the distribution function
$\Psi(\br,\bu,t) $ may be expanded with respect to the
leading moments as given  by Eqs.~(\ref{moments}).
In the parameter range  well below the  homogeneous I--N transition, where aster formation
is observed in vitro and
on which we concentrate in this work,
the first two moments are sufficient and the distribution function
may be represented in terms of $\rho(\br,t)$  and $\bt(\br,t)$ (for more details we refer to appendix \ref{appmom})
\beq
\Psi(\br,\bu,t)=\frac{1}{2\pi}\big(\rho(\br,t)+2\bu\cd\bt(\br,t)\big)~.
\eeq
To get rid of the integral kernel in Eq.~(\ref{kernel}),
a gradient expansion can be performed
as described in more detail in
appendix \ref{appgrad}.
After a rescaling of variables via
\beq
t'&=&\frac{\Dp}{l^2}t \enspace,\enspace
x'=\frac{1}{l}x \enspace,\enspace
\rho'=v_0\rho \enspace,\enspace
\bt'=v_0\bt \enspace,\enspace\nonumber\\
\alpha'&=&\frac{l}{\Dp}\alpha\enspace,\enspace
D_r'=\frac{l^2}{\Dp}D_r\enspace,\enspace
d=\frac{\Ds}{\Dp}=\frac{1}{2}\enspace,
\eeq
where for the last relation we refer to \cite{DoiEd},
we obtain in two spatial dimensions
the following coupled equations
for the density and the two components of the
polar orientation field $t_i ~(i=x,y)$ of the filaments
\begin{subequations}
\label{fund}
\beq
\label{funeq1}
\p_t\rho&=&\frac{3}{4}\Delta\rho
+\left[\frac{3}{2\pi}-\frac{\alpha}{24}\right]\n\cd\left(\rho\n\rho\right)\nonumber\\
&-&\frac{\alpha}{48}\p_i\bigg[t_i\p_j t_j+t_j\p_i t_j+t_j\p_j t_i\bigg]\nonumber\\
&-&\frac{\alpha}{C_1}\bigg\{38\n\cd\left(\rho\n\Delta\rho\right)+11\p_i\left(t_j \p_i\Delta t_j\right)\nonumber\\
& +&16\p_i\bigg[t_i\Delta\partial_j t_j+2t_j\partial_j\p_i\partial_l t_l+t_j\partial_j\Delta t_i\bigg]\bigg\},\enspace
\eeq
\beq
\p_t t_i&=&-D_r t_i+\frac{5}{8}\Delta t_i+\frac{1}{4}\p_i\nabla\cd\bt
\label{funeq2}\nonumber\\
&+&\frac{1}{4\pi}\bigg[5\p_j\left(t_i\p_j\rho\right)+\p_j(t_j\p_i\rho)+\p_i(t_j\p_j\rho)\bigg]\nonumber\\
&-&\frac{\al}{96}\p_j\bigg[3t_i\p_j\rho+t_j\p_i\rho+\rho(\p_i t_j+\p_j t_i)\bigg]\nonumber\\
&-&\frac{\al}{96}\p_i\bigg[t_l\partial_l\rho+\rho\partial_l t_l\bigg]-\frac{16\al}{C_2}\p_i\bigg[\rho\Delta\partial_l t_l+t_l\partial_l\Delta\rho\bigg]\nonumber\\
&-&\frac{\al}{C_2}\p_j\bigg[\rho\bigg(11\p_j\Delta t_i+16\p_i\Delta t_j+32\p_j\p_i\partial_l t_l\bigg)\nonumber\\
& +&16t_j\p_i\Delta\rho+32t_l\partial_l\p_i\p_j\rho+44t_i\p_j\Delta\rho\bigg]\nonumber\\
& +&\frac{1}{48}\left[\frac{\gamma}{4}-\frac{4}{\pi}D_r\right]\left\{t_j\p_j\p_i\rho-\frac{1}{2}t_i\Delta\rho\right\}
\eeq
\end{subequations}
with $C_1=23040$ and $C_2=2C_1$.

The contributions to Eqs.~(\ref{fund}) that include also   fourth order derivatives
are crucial for the determination of the instabilities evolving from
the homogeneous filament distribution  and the nonlinear
patterns in the following sections. Their importance was pointed out
recently in a comment \cite{Ziebert:2004.2}
on a previous work \cite{Liverpool:2003.2}.
It should be noted that fourth order derivatives proportional
to the translational diffusion have been neglected here because they
are overcompensated by the fourth derivatives resulting from the active current
and therefore influence the presented results
only quantitatively by a small amount (but considerably complicate the equations).

The instabilities of the homogeneous
filament distribution caused by the $\alpha$-contribution
to the active current, c.f. (\ref{vel},\ref{omegel}), and the nonlinear solutions
are determined in the following sections.
The active current proportional to $\beta$ leads to additional effects,
which are responsible e.g.
for propagating modes in the system and for the breaking of the unexpected $\bt\rightarrow-\bt$
symmetry of Eqs.~(\ref{fund}). To remind the reader, we have allowed $\bt$ to become
nonzero, so we allowed the breaking
of the $\pm\bu$-symmetry of Eq.~(\ref{conserved}), but our model equations (\ref{fund}) nevertheless have a
$\pm\bt$-symmetry, since terms like $\bt^2$ appear in the equation for $\bt$ only due to
the $\beta$-contribution of the current.
These effects are determined elsewhere \cite{Ziebert:2004.3} and
the corresponding contributions to Eqs.~(\ref{fund}) are not shown for the sake of brevity.
The active rotational current proportional to $\gamma$ is nonlinear and therefore
influences only the nonlinear behavior of pattern selection beyond threshold
as shown briefly in Sec.~\ref{weaknonlin}.

\section{Threshold for density and orientational instabilities}
\label{thresh}

Having defined the model, the first issue
in the field of pattern formation
is the determination of possible instabilities.
As calculated in this section by a linear stability
analysis, in a certain parameter range
the homogeneous basic state, which consists here in a constant filament density $\rho_0$
and a vanishing polar orientation  $\bt=0$, becomes unstable
with respect to inhomogeneous perturbations $\tilde \rho(\br,t)$
and $\bt(\br,t)$.
For this purpose, by the ansatz
$\rho (\br,t)   =\rho_0 +\tilde \rho(\br,t) $ we separate
the constant part $\rho_0$ of the filament density from
the spatially  inhomogeneous
one, $\tilde \rho(\br,t)$, and
linearize Eqs.~(\ref{fund})
with respect to  small inhomogeneous contributions
 $\tilde \rho(\br,t)$  and  $\bt(\br,t)$.
Accordingly a set of three coupled linear equations
is obtained
\beq
\label{lineq}
\p_t {\bf w} ({\bf r},t)
&=&
\mathcal{L}_0  {\bf w} ({\bf r},t)
=
\begin{pmatrix}
\mathcal{L}_{11}^{(0)} & 0 & 0\\
0 & \mathcal{L}_{22}^{(0)} & \mathcal{L}_{23}^{(0)}\\
0 & \mathcal{L}_{32}^{(0)} & \mathcal{L}_{33}^{(0)}\\
\end{pmatrix}
 {\bf w} ({\bf r},t)
~,\enspace\enspace
\eeq
for the three components of the vector
\beq\label{vdef}
{\bf w}({\bf r},t)
= \left( \begin{array}{c} \tilde \rho \\
t_x \\ t_y   \end{array}
 \right) ({\bf r},t) \,.
\eeq
The components of the linear operator ${\cal L}_0$ are
given by the expressions
\beq\label{L0entries}
\mathcal{L}_{11}^{(0)}&=&\bigg[\frac{3}{4}\left(1+\frac{2}{\pi}\rho_0\right)-\frac{\al\rho_0}{24}\bigg]\Delta
-\frac{19\hspace{1mm}\al\rho_0}{11520}\Delta^2\,,\nonumber\\
\mathcal{L}_{22}^{(0)}&=&-D_r+\frac{5}{8}\Delta+\frac{1}{4}\p_x^2-\frac{\al\rho_0}{96}(\Delta+2\p_x^2)\nonumber\\
& &-~\frac{\al\rho_0}{46080}\left(11\Delta^2+64\Delta\p_x^2\right)\,,\nonumber\\
\mathcal{L}_{23}^{(0)}&=&\left(\frac{1}{4}-\frac{\al\rho_0}{48}\right)\p_x\p_y-
\frac{\al \rho_0}{720}\Delta\p_x\p_y\,
\eeq
and the two further
components  $\mathcal{L}_{32}^{(0)}$ and $\mathcal{L}_{33}^{(0)}$ may be obtained
by permuting
$\p_x$ and $\p_y$ in $\mathcal{L}_{23}^{(0)}$ and $\mathcal{L}_{22}^{(0)}$, respectively.
Naturally the mode ansatz
\beq
\label{mode}
 {\bf w}({\bf r},t) =
{\bf E}~
\exp\left(\sigma t+ i\bf{k}\cd\bf{r}\right)
\eeq
with the wave vector ${\bf k}=(q,p)$ and the eigenvector
${\bf E}$  solves the
linear homogeneous set of equations  (\ref{lineq}) and the solubility
condition provides a third order polynomial for
the eigenvalue $\sigma$, which factorizes
into a linear and a quadratic polynomial
describing different  types  of instabilities.
Considering moderate filament densities, for intermediate values of $\alpha$,
an orientational instability with a finite wavelength
is preferred, whereas the density mode does not couple
on the level of the linear equations.  For large  values of $\alpha$
a further mode becomes unstable which resembles
a spinodal decomposition of
the filament density  driven by the motor proteins.

\begin {figure}
\epsfxsize 8.7cm
\epsfbox {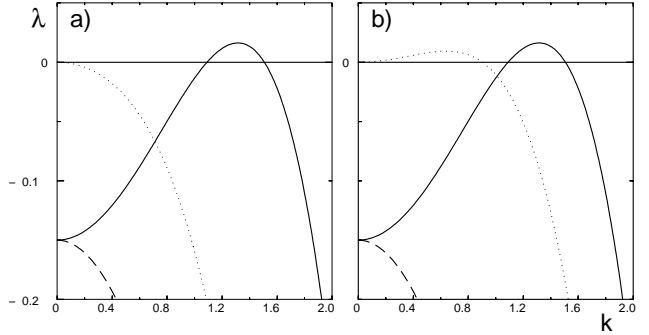}
\caption{The real parts $\lambda(k) = {\rm Re}[\sigma(k)]$
of the eigenvalue $\sigma_d$ corresponding
to the instability with
respect to density fluctuations (dotted line)
as well as the one with respect
to orientational fluctuations, $\sigma_I$ (solid line),
  are shown as a function of
the  wave number $k$ for $D_r=0.15$.
The third eigenvalue
 $\sigma_S$  is always dampened as depicted by the dashed line.
In part a) we have chosen $\alpha=21$ and therefore
$\rho_c<\rho_0<\rho_d$ leading to an orientational instability, whereas in part b) one has
 $\alpha=26$ and $\rho_d<\rho_c<\rho_0$, hence both modes
$\sigma_d$ and $\sigma_I$  have positive real parts  in a finite wave number range
(see also Fig.~\ref{rhocall}).
}
\label {dispalpha}
\end {figure}

The latter instability with respect to density fluctuations
is governed by  $\mathcal{L}_{11}^{(0)}$  and the  eigenvalue
\beq
\label{sigd}
\sigma_d = -\bigg[\frac{3}{4}\left(1+\frac{2}{\pi}\rho_0\right)-\frac{\al\rho_0}{24}\bigg]k^2
-\frac{19\hspace{1mm}\al\rho_0}{11520} ~k^4
\eeq
with  $k^2=q^2+p^2$.
The term $\propto k^4$ is always stabilizing,
but the homogeneous basic state becomes unstable
with respect to density modulations
for a positive  coefficient of $k^2$ leading to the
corresponding critical filament density
\beq
\label{critdend}
\rho_d=\frac{1}{\frac{\alpha}{18}-\frac{2}{\pi}}
\eeq
providing that
$\alpha>\frac{36}{\pi}$ holds.
The corresponding eigenvector is
${\bf E}_d = ( 1,0,0)^T$ and the dispersion is shown as the dotted lines  in Fig.~\ref{dispalpha}a) and b)
for subcritical and supercritical values respectively.

Of the remaining two eigenvalues, $\sigma_I$ and $\sigma_S$, only the first
may become positive in a finite range of $k$ as indicated by
the solid lines in Fig.~\ref{dispalpha}.
It describes an instability with respect to
orientational fluctuations with the
dispersion relation
\beq
\label{dispo}
\sigma_{I}=-D_r - \frac{1}{4}k^2\left(\frac{7}{2}-\frac{\al \rho_0}{8}+\frac{5}{768}\al \rho_0 k^2\right)\,,
\eeq
while $\sigma_S$, shown as the dashed lines in Fig.~\ref{dispalpha}, is always dampened and related to diffusion
of orientational modes.
The two corresponding   eigenvectors are
\beq
{\bf E}_I  = \left(     \begin{array}{c}  0 \\ q\\ p \end{array} \right) \,,\qquad
{\bf E}_S = \left(     \begin{array}{c}  0 \\ q\\ -p \end{array}
\right)\,.
\eeq

It should be mentioned that the eigenvalues $\sigma_d(k)$, $\sigma_I(k)$, $\sigma_S(k)$
depend only on even powers
of the wave number modulus
reflecting the rotational symmetry of the basic state $(\rho_0, {\bf t})$.
%
%
%
%
\begin {figure}
\epsfxsize 8.7cm
\epsfbox {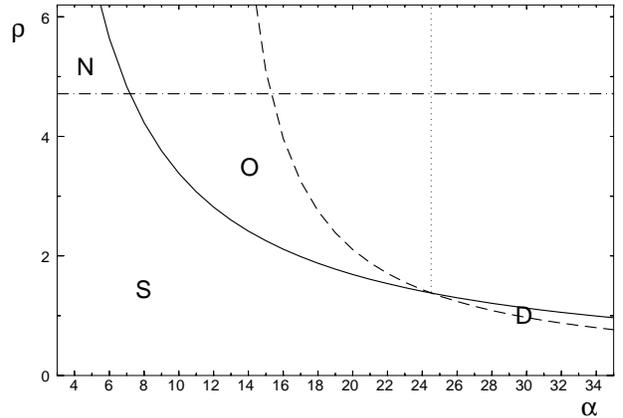}
\caption{The critical densities  $\rho_c$ (solid line)
and $\rho_d$ (dashed line) for the orientational and density instability respectively
are shown as a function of $\alpha$  and for $D_r=0.15$. The dash--dotted horizontal line
is the critical density $\rho_{IN}=\frac{3}{2}\pi$,
above which the  isotropic--nematic transition due to excluded volume interactions
takes place.
In the range referred to as $N$ one has a pure homogeneous transition to nematic order,
in range $O$ one has a motor driven spatially
periodic orientational order and in range $D$
one has modulations of the filament density.
The region $S$ denotes the parameter range where the
homogeneous solution is stable.
In the following sections of the paper we investigate the nonlinear behavior  in
the $O$-region.
}
\label {rhocall}
\end {figure}
The restabilizing $k^4$-term in Eq.~(\ref{dispo})
was missing in Ref.~\cite{Liverpool:2003.2},
as pointed out recently \cite{Ziebert:2004.2}.
However, only the interplay between the $k^2$ and $k^4$
contribution as in Eq.~(\ref{dispo}) allows
the identification of a finite wavenumber instability.
The extremal condition together with the neutral
stability condition
\beq\label{bed}
\left. \frac{d\sigma_I}{dk}\right|_{k_c}  =0\enspace,\enspace\sigma_I(k=k_c)=0~,
\eeq
allow the determination of
the critical filament density $\rho_c$ at the critical
wavenumber $k_c$, above which the orientational instability takes place:
\beq \label{rhoc}
\rho_c&=&\frac{1}{\al}\left(28+\frac{10}{3}D_r\left(1+\sqrt{1+\frac{84}{5D_r}}~\right)\right)\,,\\
k_c&=&\label{kc}4\left(\frac{12D_r}{5\al\rho_c}\right)^{1/4} \enspace.
\eeq
Fig.~\ref{dispalpha} displays the dispersions, i.e. the wavenumber dependent growth rates,
$\sigma_d(k)$, $\sigma_I(k)$ and $\sigma_S(k)$
as dotted, solid and dashed lines respectively.
In Fig.~\ref{rhocall}, the critical density  $\rho_d$ (dashed line) for an instability
with respect to inhomogeneous
density fluctuations and $\rho_c$ (solid line)
with respect to inhomogeneous orientational fluctuations ${\bf t}({\bf r},t)$
are shown as a function of motor activity $\alpha$.
The dash-dotted line describes the critical density  $\rho_{IN}$ (\ref{rhoIN}) above which
the homogeneous isotropic--nematic transition takes place.
On the left side of the dotted line, orientational fluctuations have
lowest threshold,
while on the right side density fluctuations become unstable at first.
For increasing values of the rotational diffusion coefficient $D_r$, which determines how negative
the dispersions $\sigma_I(k)$ and $\sigma_S(k)$ start from $k=0$,
the solid line is shifted upwards in  Fig.~\ref{rhocall}
decreasing the $\alpha$-range wherein the
orientational instability has lowest threshold with $\rho_c<\rho_{IN}$.

Beyond some critical value of $\beta$, the $\beta$-contribution to the active current (\ref{acurrent},\ref{vel})
induces an oscillatory bifurcation from the homogeneous filament distribution
as described in a forthcoming work, where also
the nonlinear behavior of the oscillatory patterns is
discussed \cite{Ziebert:2004.3}.  The contribution of the rotational current
(\ref{arcurrent},\ref{omegel}) to Eqs.~(\ref{fund}) includes only nonlinear
terms and therefore does not influence the thresholds
of the instabilities. Its influence on the nonlinear behavior
of stationary periodic patterns is among other things discussed in the next section.

\section{Nonlinear analysis of the orientational instability}
\label{weaknonlin}
The amplitude  of the linear solution in
Eq.~(\ref{mode}) is limited  by the
terms in  Eqs.~(\ref{fund})
that are nonlinear  with respect to $\tilde \rho$
and ${\bf t}$. Here we investigate the
weakly nonlinear behavior beyond the orientational instability, i.e. in the region referred to as $O$
in Fig.~\ref{rhocall} where $\rho_c<\rho_0<\rho_d$ holds.
This can be done numerically as exemplified in the next section and even
analytically if $\rho_0$ is not immoderately beyond $\rho_c$ defined in (\ref{rhoc}):
for a supercritical
bifurcation,  the amplitude of the mode initially growing with $\sigma_I(k)$
is small immediately above threshold
and may be determined in this range by a perturbative analysis, the
so-called  amplitude equation \cite{crossho:93},
whose derivation is sketched in appendix \ref{Ampderiv}.

The generic form of  amplitude equations depends on the preferred
 pattern  beyond a stationary supercritical bifurcation.
In two spatial  dimensions the pattern is
either spatially periodic in one direction (stripes), in two (squares/rectangles) or
in three directions (hexagonal patterns) \cite{crossho:93}.
Numerical simulations of the basic equations (\ref{fund}) indicate,
as described in  Sec.~\ref{numsim},
that stripe and square patterns are favored immediately above the  threshold
of the orientational instability. Moreover, hexagonal structures are not driven in this system
due to the overall
up-down symmetry of Eqs.~(\ref{fund}) with respect to $\bt$,  c.f. appendix \ref{Ampderiv} as well.

Square patterns at threshold may be
described ana\-ly\-ti\-cal\-ly
by a superposition of two linear modes (i.e. stripes) with
orthogonal wave numbers which can be chosen without restriction as ${\bf k}_1 =(k_c,0)$
and ${\bf k}_2=(0,k_c)$ with $k_c$ given by Eq.~(\ref{kc}) leading to the ansatz
\beq\label{ansatz}
{\bf w}_1 =
\begin{pmatrix}\rho_1 \\t_{1x}\\t_{1y}\\ \end{pmatrix}=\begin{pmatrix}0\\X\\0\\ \end{pmatrix}e^{ik_c x}
+\begin{pmatrix}0\\0\\Y\\ \end{pmatrix}e^{ik_c y}+c.c.\,,
\eeq
where $c.c.$ means the complex conjugate.
In a small neighborhood of the threshold $\rho_c$, measured by
the dimensionless control parameter
\beq\label{rhoexpand}
\varepsilon = \frac{\rho_0-\rho_c}{\rho_c}~,
\eeq
the time-dependent amplitudes $X(t)$ and $Y(t)$
 of  the  square
pattern  are determined by  two generic coupled nonlinear
equations, the so-called amplitude equations \cite{Segel:65.1,crossho:93}
\begin{subequations}
\label{Aeq}
\beq \label{Aeq1}
\tau_0\p_t X&=&\varepsilon X -g_1|X|^2 X-g_2|Y|^2 X\enspace,\\
\label{Aeq2}
 \tau_0\p_t Y&=&\varepsilon Y -g_1|Y|^2 Y-g_2|X|^2 Y\enspace.
\eeq
\end{subequations}
The coefficients $\tau_0$, $g_1$  and $g_2$ are derived
in  appendix \ref{Ampderiv}.
The specific system under consideration enters into the description only in these coefficients via the
parameters of the basic equations (\ref{fund}), namely $\alpha$, $D_r$ and
$\gamma$, while Eqs.~(\ref{Aeq}) are universal for a whole symmetry class of
pattern forming systems.

Apart from the trivial solution $X_0=Y_0=0$, reflecting the stable
homogeneous system,
the coupled amplitude equations $(\ref{Aeq})$
 have also stationary finite amplitude solutions. These are
 at first
\begin{subequations}\label{rollsol}
\beq
 X_0=\pm\sqrt{\frac{\varepsilon}{g_1}}~,~Y_0=0~\,,\\
 X_0=0~,~Y_0=\pm\sqrt{\frac{\varepsilon}{g_1}}~\,,
\eeq
\end{subequations}
which correspond according to Eq.~(\ref{ansatz})
to stripes periodic  either  in $x$- or in $y$-direction.
Secondly, there is the stationary solution of equal amplitudes
\beq\label{squaresol}
X_0=Y_0=\pm \sqrt{\frac{\varepsilon}{g_1+g_2}}~\,,
\eeq
which constitutes a square pattern. In real space, this square pattern
in terms of the components of the vector field ${\bf t}({\bf r},t)$ resembles the structures
which are called asters and found in numerous experiments \cite{Nedelec:1997.1} as will
become clear from the simulation pictures in Sec.~\ref{numsim}.

As summarized in appendix \ref{squarestab}, by
a linear stability analysis of the (nonlinear)
stationary solutions given by Eqs.~(\ref{rollsol}) and
(\ref{squaresol})
one finds stable stripes
as the preferred solution  in the range  $g_2> g_1>0$ of the nonlinear coefficients.
In the parameter range $|g_2| < g_1$
the square pattern is preferred \cite{Segel:65.1}.

\begin {figure}
\epsfxsize 8.7cm
\epsfbox {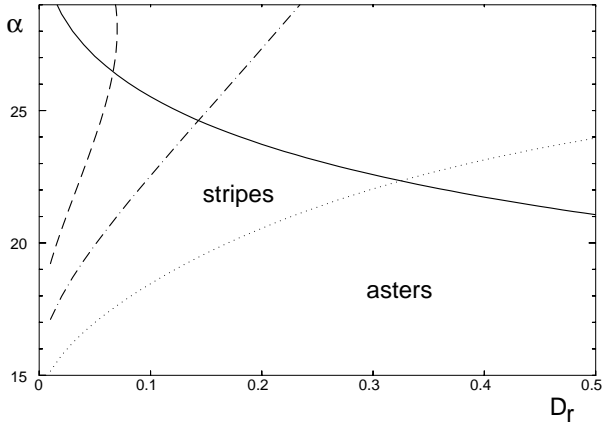}

%
\caption{The stability regions of stripes and asters are shown as
calculated by the amplitude expansion method.
The two critical densities $\rho_c$ and  $\rho_d$
coincide along the solid line and beyond it
instabilities with respect to density
modulations are preferred not included in our present nonlinear analysis.
The dotted line is given by $g_1=g_2$ and
separates the range of stable square patterns
 (asters)  from the range of
stable stripe patterns.  Along the  dash-dotted line one has
$g_1=0$ and the bifurcation to stripes changes from
supercritical (below)
to a subcritical  one (beyond).
Between  the dashed line, which is determined by $g_1=-g_2$,
and  the dash-dotted line asters can still exist but are unstable while
the amplitudes of stripes cannot be determined by our
lowest order expansion. Beyond the dashed line, also asters bifurcate subcritical.
}
\label {phdiag}
\end {figure}

These criteria for $g_1$ and $g_2$  may
be translated according to
their parameter dependence into the $D_r$-$\alpha$ plane, as
shown in Fig.~\ref{phdiag} with
the nonlinear contributions of the rotational current neglected at first, i.e.
$\gamma=0$.
The analytic calculations presented here are valid below the solid line in Fig.~\ref{phdiag},
since above the density instability takes place.
The dotted line corresponds to the condition $g_1=g_2>0$
which separates the range of stable squares from stable stripe solutions.
Along the dash--dotted line in Fig.~\ref{phdiag}
the bifurcation from the homogeneous basic state
to the stripe pattern changes its behavior from
a supercritical (below) to a subcritical one (above), so the expansion method is no more effective.

\begin {figure}
%
\epsfxsize 8.7cm
%
\epsfbox {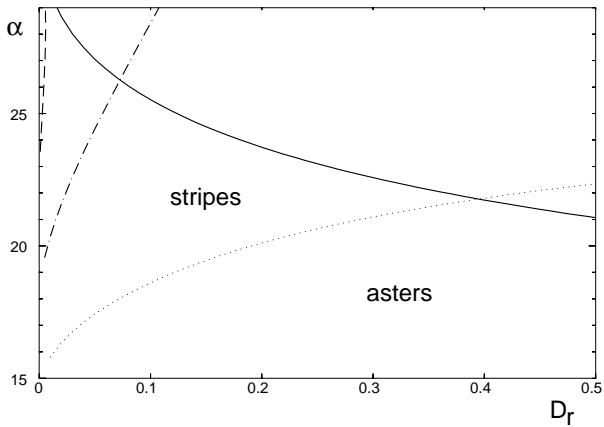}

%
\caption{ The same diagram as in Fig.~\ref{phdiag}
but here the nonlinear contributions of rotational diffusion
and active rotational currents, c.f. Eq.~(\ref{omegel}),
are taken into account with $\gamma=\alpha$.
The region of stable stripes considerably broadens
while the regions of subcritical bifurcations are moved to higher $\alpha$ and lower $D_r$ values.
}
\label {phdiaggam}
\end {figure}

Nonlinear effects of the rotational diffusion and  the active rotational
current are described by the last term in Eq.~(\ref{funeq2}). Since they
make no contribution to the linear operator, they can only change the pattern selection, i.e.
the stability regions of the nonlinear solutions.
Taking them into account, the
bifurcation behavior from the homogeneous basic state
is changed as shown in
Fig.~\ref{phdiaggam} for $\gamma=\alpha$, showing that these
effects enlarge the range of stable stripe patterns in
the $D_r$-$\alpha$ plane.

%

%

One may complement this outline of the bifurcation behavior
by a discussion of the two decisive model
pa\-ra\-me\-ters, namely $\alpha$ and $D_r$, and analytical estimates for them.
Simple models for motor proteins \cite{Howard:2001,Sekimoto:1996.1,Kruse:2000.1,Liverpool:2003.2}
imply that the rate $\alpha$ of the translational active transport grows linearly with
the active motor density $\rho_m$
and with the length of the filaments, i.e.
 $\alpha\propto\rho_m l $.
Hence this rate can be controlled by the cell in the most effective way
by the degree of motor activity (i.e. by regulating the ATP concentration)
as well as on a much larger timescale by the density of the motors and the filament length.

For the rotational and
translational diffusion coefficients in a {\it dilute solution},
calculations taking the hydrodynamic interaction into account \cite{DoiEd}
propose the analytical expressions
\beq
D_r=\frac{3\ln(l/b)}{\pi\eta l^3}\enspace,
\qquad \Dp=\frac{\ln(l/b)}{2\pi\eta l}\enspace.
\eeq
In our scaled units this means
$D_r'=\frac{l^2}{\Dp}D_r=6$, lying
far in the range of squares (asters).

For {\it semi--dilute solutions} one obtains in a similar manner
$D_r=6\Dp/(l^2(1+c_r\rho')^2)$
or
\beq
D_r'=\frac{l^2}{\Dp}D_r=\frac{6}{(1+c_r\rho')^2}
\eeq
in scaled units,
where $c_r\simeq 1$ is a geometry factor from a tube model calculation.
Since stripes or bundle-like structures are stable for $D_r'<0.3$--$0.4$, c.f. Figs.~\ref{phdiag}
and \ref{phdiaggam},
one needs a rather high (but possible) filament density for such a one-dimensional ordering.

According to these estimates
asters are the most likely pattern occurring
above a stationary bifurcation in dilute or semi-dilute two-dimensional motor--filament--systems.
For bundle-like structures to emerge rather high filament densities are needed, which is
physically intuitive from the overlap nature of all the interactions, namely excluded volume
and motor-induced filament-filament interaction.

\section{Results of  numerical simulations}
\label{numsim}

Besides the weakly
nonlinear analysis described in the previous section,
the basic equations (\ref{fund}) have been solved
numerically in order to check the
validity range of the perturbation analysis
and further explore the solution space.
For this purpose a Fourier Galerkin pseudo-spec\-tral me\-thod
has been used
imposing periodic boundary conditions on the system.

\begin {figure}
\epsfxsize 8.7cm
\epsfbox {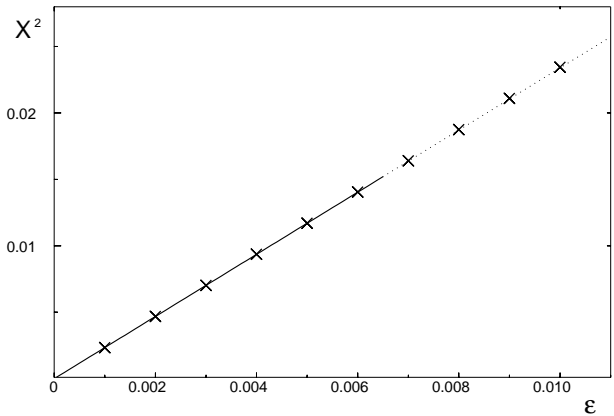}
\caption{A comparison of the  amplitude of a stripe solution
as obtained  in simulations (crosses) with the analytical calculation
$X_0 =\pm (\varepsilon/g_1)^{1/2}$ as given by Eq.~(\ref{rollsol}) is shown.
The agreement remains well for higher values of $\varepsilon$,
but in the range between $\varepsilon\simeq0.006-0.007$
a secondary bifurcation takes place rendering the high-$\varepsilon$ solutions unstable.
Parameters are $D_r=0.5$, $\alpha=21$ and $\gamma=0$.
}
\label {vglAeps}
\end {figure}

Since the validity range
of the amplitude expansion with respect to the
control parameter $\varepsilon$
is not known a priori, in Fig.~\ref{vglAeps} we compare
the amplitude of a stripe solution as obtained by a numerical solution of the  basic equations
(\ref{fund})
with the analytical results given by  Eq.~(\ref{rollsol}).
Close to threshold there is   nearly perfect agreement
between both approaches.
However the validity range of the amplitude equations
is actually restricted to a range below  $\varepsilon \sim 0.006$ for the
parameters used in Fig.~\ref{vglAeps}.
Around this value a secondary instability takes place, which is not
taken into account in the perturbation expansion.
The numerical simulations show that beyond this secondary
instability a pronounced accumulation of the filaments
to densities even higher than $\rho_{IN}$ appear
accompanied with high alternating orientations.
These solutions are numerically stable but nevertheless in an invalid
range of our model since the
nematic order parameter, c.f.
Eqs.~(\ref{moments}), has been neglected in the moment expansion which is not legitimate anymore
 -- so we reject showing pictures here.

%

%
\begin {figure}
%
\epsfxsize 8.7cm
\epsfbox {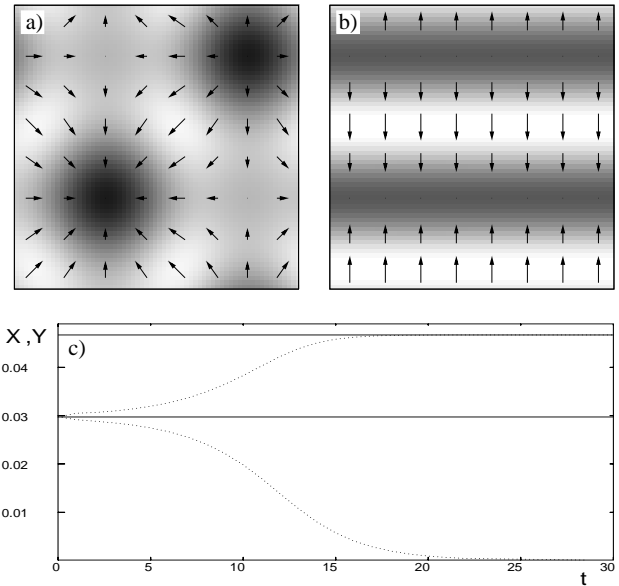}

%
\caption{ A scenario is shown
where a repetitive structure of asters as displayed in (a)
is slightly disturbed
at a point in parameter space where stripes are preferred, namely at
$\alpha=21$ and
$D_r=0.15$ ( and $\varepsilon=5\cdot10^{-4}$ ).
In part (c)  the
solid lines correspond to the
analytical values of the amplitudes $X_s=Y_s=0.0297$ for squares and $X_r=0.0467$ for
stripes respectively while the dotted lines show how the
amplitudes $X$ and $Y$ of the polar orientation components  $t_x$ and $t_y$
evolve in time from the unstable square pattern to the stationary stripe solution in (b).
}
\label {str}
\end {figure}

In addition to the validity range with respect to
$\varepsilon$, one may also confirm
numerically the stability of the nonlinear solutions as predicted
in Fig.~\ref{phdiag} by the weakly nonlinear analysis.
As an example,
we start with a square solution as shown in  Fig.~\ref{str}a)
at the point
$\alpha=21$ and
$D_r=0.15$ in parameter space (and $\varepsilon=5\cdot10^{-4}$)
belonging according to Fig.~\ref{phdiag}
to the region of stable stripe patterns. After a slight perturbation,
the simulated temporal evolution in Fig.~\ref{str}c) shows that only one of
the initially equal amplitudes remains finite in the long time limit leading to the predicted
stationary stripe pattern displayed in Fig.~\ref{str}b).
By several numerical runs we confirmed the analytically predicted stability diagrams
in Figs.~\ref{phdiag} and \ref{phdiaggam}.

%
\begin{figure}[hbt]
\begin{center}
\epsfxsize 6cm
\epsfbox {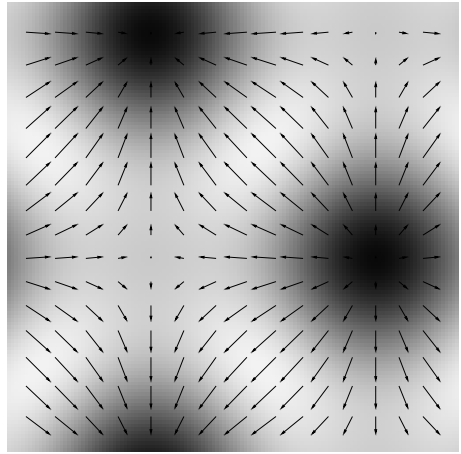}
\caption{A simulation of Eqs.~(\ref{fund}) in the range of stable stationary asters
is shown as
a superposed plot of the orientation field (arrows) and the filament density
(dark color coding low density, light color high density).
Parameters are $\alpha=21$,
$D_r=0.5$, $\varepsilon=5\cdot10^{-5}$.
}
\label {densut}
\end{center}
\end {figure}

In the parameter range of stable asters,
a vectorplot of the orientation field superposed on the color coded filament density is presented
in Fig.~\ref{densut}.
The filament orientation is indicated by arrows,
the length being a measure of the degree of orientation.
The density is high in the bright regions and low in the dark ones.
At the right hand side in Fig.~\ref{densut}, one can spot an aster with arrows pointing radially from a
center with lowered filament density opposing to an inverse aster top left with
arrows pointing radially into the center (we remind the reader that periodic boundary conditions are imposed).
The centers of the asters have lowered filament densities
which can be explained by the nonlinear analysis: in the derivation of the amplitude equations
in appendix \ref{Ampderiv} one can clearly see that
the growing amplitudes of the orientation modulations exite higher density modes
- which then limit the orientation amplitudes to render the system stable -
and therefore in the center of an aster, where the orientation vanishes,
there is no need for a high density.
Both density and the degree of orientation reach their maximum in between the asters and
two saddle-like structures building up a square with the two opposing aster centers
complete the repetitive structure of the pattern at threshold in the
motor--filament--system.

\section{Summary and conclusions}
\label{sumconc}

For a model of interacting biopolymers and motor proteins,
a nonlinear competition between the formation of
stripes and square patterns has been described for the first time.
The model is based on a nonlinear (and nonlocal) Smoluchowski equation for the distribution
function of rigid rods
extended by active currents to account for the motor-mediated relative motion of the filaments.
This equation can be approximated by a local equation
of motion under the  assumption
that the spatial variations are small on the length scale of the filaments, which
allows a
gradient expansion \cite{Liverpool:2003.2}. However, this
expansion  has to be continued up to fourth order in the gradients as
pointed out recently in order to reasonably evaluate the linear
stability of the homogeneous  filament distribution
\cite{Ziebert:2004.2}. This  expansion as well as the linear
stability of the basic state are described comprehensively.
The homogeneous state may become unstable  with respect
to  density, orientational and nematic instabilities, but in different
parameter ranges as shown in Fig.~\ref{rhocall}.
The two crucial model parameters are the rate of active translational transport
and the rotational diffusion coefficient which both can be regulated by the cell
by creating fi\-la\-ments, motor proteins and the fuel ATP as discussed and estimated
at the end of Sec.~\ref{weaknonlin}.

Our subsequent nonlinear analysis focuses
on the orientational instability, which may be of high
biological relevance for aster-like structures as in the mitotic spindle which have
also be seen in in vitro systems recently \cite{Nedelec:1997.1,Smith:04.1}.
This analysis is to a large extent  analytically in terms
of an  amplitude
expansion technique yielding two coupled equations, namely Eqs.~(\ref{Aeq}), for the
amplitudes of the spatial modulation of the orientational field in $x$- and $y$-direction.
These two coupled equations
determine the amplitudes of square (asters) and stripe patterns and
they provide also the stability boundary between both
solutions near threshold. Simulations with a spectral code and periodic boundary conditions
confirmed these analytical results.

Estimates for the two model parameters $\alpha$ and $D_r$ of the
basic equations (\ref{fund}), c.f. end of Sec.~\ref{weaknonlin},  suggest that
asters are the most likely pattern occurring
above a stationary bifurcation in dilute or semi-dilute two-di\-men\-sio\-nal motor--filament--systems.
Stripes may be expected only for rather high filament densities
which can be traced back to the overlap nature of both present
interactions, namely excluded volume
and motor-induced fi\-la\-ment-fi\-la\-ment interaction.
Once again both the type of the pattern, i.e. asters or stripes,
as well as its wavelength could be governed by the cell
by creating fi\-la\-ments, motor proteins and ATP.

Experiments as described in Refs.~\cite{Urrutia:1991.1,Nedelec:1997.1,Surrey:98.1}
show that in a model system comprised of microtubules and a single type of motor
in a confined  geometry
asters may arise but are unstable to a vortex
with motors rotating around the center, whereas in unconfined geometries, with increasing
motor concentration, the arising patterns are vortices, mixtures of asters and vortices, asters
and finally bundles of microtubules for high motor densities.
This is in agreement with our calculations visualized in Fig.~\ref{phdiag},
as for moderate values of $D_r$ the asters in our model become
unstable against stripe formation for increasing $\alpha$, which is
proportional to the (homogeneous) density of motors.
Since we only considered periodic boundary conditions,
we did not find vortices, which nevertheless
may be possible if a confined geometry is simulated.
In contrast to this good agreement with microtubule-motor systems,
recent experiments on actin-myosin systems \cite{Smith:04.1} showed that
the models at hand can not
be directly used in that case. But modifying the present model to account for ATP depletion
and stable cross-linking of the filaments by inactive myosin complexes, it is even able to explain
the different aster formation mechanism in actin \cite{Smith:04.1,Ziebert:2004.3}.

The first attempt of describing patterns in micro\-tu\-bu\-les and motors in Ref.~\cite{Nedelec:1997.1}
was done by a molecular dynamics simulation where asters and vortices
could be reproduced. In a subsequent publication \cite{Leibler:2001.1}, the case of two types of motors
walking on the filaments in opposite directions was investigated and a pattern referred to as a
network of interconnected poles was obtained which is very similar to
the pattern near threshold displayed in Fig.~\ref{densut}.
A first macroscopic description of the problem was given in Ref.~\cite{Bassetti:2000.1},
where coarse-grained equations derived from a spin-like model displayed stripe patterns.
Finally, in Ref.~\cite{Kardar:2001.1} equations
for the motor density and a phenomenological vector field describing the polar orientation
were proposed,
which reproduced asters and, in a confined geometry, vortices, but were not able to describe stripe patterns.
This can be traced back to the omission of an explicit equation
for the filament density. More severely, as will be worked out in Ref.~\cite{Ziebert:2004.3},
in the model presented here
linear couplings between the density and the orientation field via the active transport terms
proportional to $\beta$ lead to
an oscillatory instability and propagating modes through the motor--filament--system, which cannot be
described by Ref.~\cite{Kardar:2001.1} but should be highly relevant for vital processes like cell spreading on a
substrate as measured for instance in Ref.~\cite{Doeb:04.01}.

Some experiments indicate  that the spatially inhomogeneous
distribution of the motor density may become a further
relevant degree of freedom  \cite{Nedelec:2001.2} and therefore
should be taken into account for various purposes in future models.
This degree of freedom in combination with stable crosslinking may also give rise to a different instability
in actomyosin \cite{Smith:04.1,Ziebert:2004.3}.
From the theoretical point of view it would be interesting to extend the moment expansion scheme
one step further yielding an equation for the
nematic order parameter tensor. This would allow to take also
the effects of nematic order into account and to study its interplay
with the polar orientation in the presence of motors, which may lead
to an interesting model system of a new symmetry class.

%
\appendix
%
\section{Symmetries of the motor-mediated velocities}\label{symmvel}
The explicit form of the motor-mediated translational and angular velocities,
namely Eqs.~(\ref{vel}) and (\ref{omegel}), can be obtained by writing down the
simplest terms fulfilling the conservation laws and symmetries of the system.

If one considers an interacting filament pair in the absence of external forces and torques, both
momentum and angular momentum of the pair have to be conserved. Hence the motor-mediated translational
and angular velocities,
$\mathbf{v}=\dot\br-\dot\br'$ and $\bm{\omega}=\dot\bu-\dot\bu'$,
have to be odd under the transformation $(\br,\bu;\br',\bu')\rightarrow(\br',\bu';\br,\bu)$.
Translational invariance leads to $f(\br,\br')=f(\br-\br')$ for $f=\mathbf{v},\bm{\omega}$.
Together with the rotational invariance, $\mathbf{v}$ has to be odd and $\bm{\omega}$ even
under the transformation $(\br-\br';\bu,\bu')\rightarrow(\br'-\br;-\bu,-\bu')$.
The simplest terms fulfilling the above conditions for $\mathbf{v}$ are proportional to $\br-\br'$ or $\bu-\bu'$
and for $\bm{\omega}$ proportional to $\mathbf{u}\times\mathbf{u'}$. One can thus write
down \cite{Liverpool:2003.2}
\beq
\mathbf{v}(\mathbf{r\hspace{-1mm}-\hspace{-1mm}r'}\hspace{-1mm},
\mathbf{u},\mathbf{u'})&=&\frac{\alpha}{2}\frac{\mathbf{r-r'}}{L}
\frac{1+\mathbf{u}\cdot\mathbf{u'}}{|\mathbf{u}\times\mathbf{u'}|}+
\frac{\beta}{2}\frac{\mathbf{u'-u}}{|\mathbf{u}\times\mathbf{u'}|}\enspace,\\
\bm{\omega}(\mathbf{u},\mathbf{u'})&=&\gamma(\mathbf{u}\cdot\mathbf{u'})
\frac{\mathbf{u}\times\mathbf{u'}}{|\mathbf{u}\times\mathbf{u'}|}\enspace.
\eeq
The term proportional to $\alpha$ is made orientation-dependent (it could be generalized by allowing
two different prefactors) and the plus sign favours parallel alignment. In the term
proportional to $\gamma$, spatial dependence has been neglected and the prefactor $\bu\cdot\bu'$ models the tendency of motors to bind
on two filaments which share an angle smaller than $\frac{\pi}{2}$.
The common factor $|\mathbf{u}\times\mathbf{u'}|^{-1}$ is just a normalization
to the interaction volume.

%
\section{The gradient expansion of the interaction integrals}\label{appgrad}
The  excluded volume interactions, c.f. Eq.~(\ref{Vex}),  as well as the
motor induced rod--rod interaction, c.f. Eqs.~(\ref{acurrent}), are
defined by overlap integrals. Hence the equation of motion (\ref{conserved})
for the probability distribution function $\Psi(\mathbf{r},\mathbf{u})$
is nonlocal and its solution is exceedingly difficult.
Assuming that spatial variations are small on the length scale of a filament,
 in order to deal with a
local equation one can perform a systematic
expansion  of the integrals
with respect to gradients of the probability distribution function,
as described in this appendix.

The interaction kernel given by  Eq.~(\ref{kernel})
can be  expressed in terms of Straley coordinates \cite{Straley:1973.2}, which in two dimensions are defined by
\beq
\br-\br'=\bu\zeta+\bu'\eta \,,
\eeq
with the parameter constraint
$-L/2 < \zeta,\eta < L/2$ (L the filament length)
and the Jacobian  $|\bu\times\bu'|$.
Thus the excluded volume interaction, namely Eq.~(\ref{Vex}),
is determined by the  four dimensional integral
\beq
V_{ex}(\mathbf{r},\mathbf{u})&=&\int\hspace{-1mm}d\mathbf{u'} \int d\mathbf{r'}
\int\limits_{-L/2}^{L/2}d\zeta\int\limits_{-L/2}^{L/2}d\eta  ~|\bu\times\bu'| \nonumber\\
& &\quad\quad\delta\left(\br-\br'+\bu\zeta+\bu'\eta\right)\hspace{1mm}\Psi(\mathbf{r'},\mathbf{u'})~,
\eeq
which  may be approximated by a Taylor series expansion
of the  $\delta$--function with respect to $\eta$ and $\zeta$.\footnote{Equivalently, after
performing the $\mathbf{r}'$--integration the shifted distribution function
$\Psi\left(\br+\bu\zeta+\bu'\eta\right)$ may be expanded.} Performing the
$\eta$ and $\zeta$  integration one finally obtains
\beq
V_{ex}(\mathbf{r},\mathbf{u})&=&L^2\int\hspace{-1mm}d\mathbf{u'}|\bu\times\bu'|\cdot\nonumber\\
& &\hspace{-8mm}\left[1+\frac{L^2}{24}\left\{\left(\mathbf{u}\cdot\partial_{\mathbf{r}}\right)^2+\left(\mathbf{u'}\cdot\partial_{\mathbf{r}}\right)^2\right\}\right]\Psi(\mathbf{r},\mathbf{u'})\,,
\eeq
where terms up to second order in the spatial derivatives have been taken into account.
The prefactor $L^2$ reflects the two-dimensional excluded volume.

By a similar procedure, the contribution to the active current proportional to
$\alpha$ can be evaluated as
\beq
J^a_t&=&-\frac{\alpha L^3}{24}\partial_{i}
\int d\mathbf{u'}\Psi(\mathbf{r},\mathbf{u})\left(1+\mathbf{u}\cdot\mathbf{u'}\right)\cdot\nonumber\\
& &\hspace{-0.8cm}\bigg\{u_{i}\hspace{-0.1cm}\left(\mathbf{u}\cdot\partial_{\br}\right)
\hspace{-0.1cm}\left[1-\frac{L^2}{16}
\hspace{-0.1cm}\left(\frac{2}{5}\left(\mathbf{u}\cdot\partial_{\br}\right)^{2}
\hspace{-0.1cm}+\hspace{-0.1cm}\frac{2}{3}\left(\mathbf{u'}\hspace{-1mm}\cdot\partial_{\br}\right)^{2}\right)\right]\nonumber\\
& &\hspace{-0.8cm}+u'_{i}\hspace{-0.1cm}\left(\mathbf{u'}\hspace{-1mm}\cdot\partial_{\br}\right)
\hspace{-0.1cm}\left[1-\frac{L^2}{16}
\hspace{-0.1cm}\left(\frac{2}{5}\left(\mathbf{u'}\hspace{-1mm}\cdot\partial_{\br}\right)^{2}
\hspace{-0.1cm}+\hspace{-0.1cm}\frac{2}{3}\left(\mathbf{u}\cdot\partial_{\br}\right)^{2}\right)\right]\hspace{-0.1cm}\bigg\}
\Psi(\mathbf{r},\mathbf{u'}).\nonumber\\
\eeq
One should note that due
to the $\zeta,\eta$-integrations,
odd powers of $\partial_{\mathbf{r}}$ vanish in both expressions,
which is the mathematical reason underlying the rotational symmetry of Eqs.~(\ref{fund}).

\section{The moment expansion method}\label{appmom}

The method for deriving  the coupled Eqs.~(\ref{fund})
for the density and the orientation field of the filaments from the
underlying Smoluchowski-equation, Eq.~(\ref{conserved}),
is similar to the calculations presented in Ref.~\cite{Doi:88.1}, where it has been used to obtain
equations describing the initial stage of spinodal decomposition during the isotropic--nematic transition
of a three-dimensional hard-rod fluid.
In both cases, the starting point is an approximate description
of the probability distribution function $\Psi(\mathbf{r},\mathbf{u},t)$
by its moments, but in contrast to a usual liquid crystal as considered in Ref.~\cite{Doi:88.1},
now the rods are polar with respect to the
motor action. Thus the $\pm\bu$-symmetry is broken
and the first moment, namely the filament orientation field ${\mathbf t}$, does not
vanish here and cannot be omitted. Hence one has to approximate $\Psi(\mathbf{r},\mathbf{u},t)$
by its first two moments,
\beq\label{momexp}
\Psi(\mathbf{r},\mathbf{u},t)\simeq\frac{1}{2\pi}\big(\rho(\mathbf{r},t)+2\mathbf{u}
\cdot\mathbf{t}(\mathbf{r},t)\big)\enspace,
\eeq
with the filament density $\rho(\mathbf{r},t)$ and the orientation
field $\mathbf{t}(\mathbf{r},t)$, c.f. Eqs.~(\ref{moments}).

The validity of the representation in Eq.~(\ref{momexp}) can be seen immediately by using it to evaluate the
first moments which correctly yields
$\int d\bu~\Psi(\mathbf{r},\mathbf{u})  =\rho(\mathbf{r})$ and
$\int d\bu\hspace{0.5mm}\bu\hspace{0.5mm}\Psi(\mathbf{r},\mathbf{u})=\bt(\mathbf{r})$.
Inserting  the currents
as defined by Eqs.~(\ref{current}) and (\ref{acurrent}) as well as
applying the gradient expansion given in appendix~\ref{appgrad}, then an
 integration of  Eq.~(\ref{conserved})
by $\int d\mathbf{u}$ and
$\int d\mathbf{u}\hspace{0.3mm}\mathbf{u}$
 yields two evolution equations for $\rho(\mathbf{r},t)$ and $\mathbf{t}(\mathbf{r},t)$ respectively,
as given by Eqs.~(\ref{fund}).

The remaining integrals involved are merely orientational averages. If one defines
\beq\label{mean_def}
\langle A(\bu)\rangle_{\bu}=\int\frac{d\bu}{2\pi}A(\bu)=\int_0^{2\pi}\frac{d\theta}{2\pi}A(\theta)\enspace,
\eeq
where $\theta$ parameterizes the unit vector $\bu$ in two dimensions, the following
formulas are useful and easily proven:
All mean values depending on
odd powers of $\bu$ vanish as well as the mean values of
any product of $|\bu\times\bu'|$ and odd powers of $\bu$ or $\bu'$,
since $|\bu\times\bu'|=\sqrt{1-(\bu\cdot\bu')^2}$ contains only even powers of $\bu,\bu'$.
For the even powers one gets
\beq\label{isomean1}
\langle u_{\alpha}u_{\beta}\rangle_{\bu}=\frac{1}{2}\delta_{\alpha\beta}\enspace,
\eeq
\beq
\langle u_{\alpha}u_{\beta}u_{\mu}u_{\nu}\rangle_{\bu}=\frac{1}{8}\left(
\delta_{\alpha\beta}\delta_{\mu\nu}+\delta_{\alpha\mu}\delta_{\beta\nu}+\delta_{\alpha\nu}\delta_{\beta\mu} \right)
\eeq
and
\beq
\langle u_{\alpha}u_{\beta}u_{\mu}u_{\nu}u_{\sigma}u_{\tau}\rangle_{\bu}
=\frac{1}{48}\sum_{\mathcal{P}}
\delta_{ij}\delta_{kl}\delta_{mn}\enspace,
\eeq
where $\mathcal{P}$ means all the permutations
of $\{\alpha,\beta,\mu,\nu,\sigma,\tau\}$
generating different index combinations of the Kronecker delta
product. Furthermore one needs
\beq\label{isomean2}
\langle|\bu\times\bu'|\rangle_{\bu'}=\int \frac{d\theta'}{2\pi}|\sin\left(\theta-\theta'\right)|=\frac{2}{\pi}
\eeq
and
\beq
\langle|\bu\times\bu'|u'_{\alpha}u'_{\beta}\rangle_{\bu'}=-\frac{2}{3\pi}\left(
u_{\alpha}u_{\beta}-2\delta_{\alpha\beta}\right)\enspace.
\eeq
The operator of rotational diffusion in two dimensions as given by Eq.~(\ref{rotop}) can be expressed as
\beq
[\mathcal{R}]_i=[\mathbf{u}\times\partial_{\mathbf{u}}]_i=\delta_{i3}\left(u_1u_2'-u_2u_1'\right)
\eeq
and for the rotational terms, the following partial integration formula
\beq
\langle A(\bu)\mathcal{R}\left[B(\bu)\right]\rangle_{\bu}=-\langle\mathcal{R}\left[A(\bu)\right]B(\bu)\rangle_{\bu}
\eeq
is crucial to simplify calculations.

As an example, we calculate a rather simple term, namely the excluded volume contribution
to the equation of motion for the filament density $\rho$, c.f. (\ref{funeq1}), which
is  proportional to $D_{\parallel}$ and second order in
spatial derivatives.
So we have to deal with
\beq
D_{\parallel}\hspace{-1mm}\int\hspace{-1mm} d\bu\enspace u_iu_j\partial_i\left(\Psi(\bu)\partial_j L^2\hspace{-1mm}\int\hspace{-1mm} d\bu'|\bu\times\bu'|\Psi(\bu')\right),
\eeq
where spatial coordinates have been suppressed.
Substituting the representation Eq.~(\ref{momexp}) twice and using the nomenclature introduced in Eq.~(\ref{mean_def}), we have to calculate
\beq
\langle u_iu_j\partial_i\left(\left[\rho+2u_kt_k\right]\partial_j \langle|\bu\times\bu'|\left[\rho+2u'_lt_l\right]\right)\rangle_{\bu}\rangle_{\bu'}.
\eeq
The contributions $u_kt_k$, $u'_lt_l$ vanish because of their oddness. Thus we are left with
\beq
\langle u_iu_j\partial_i\left(\rho\partial_j \langle|\bu\times\bu'|\rangle_{\bu'}\rho\right)\rangle_{\bu}
\eeq
which can be simplified to
\beq
\frac{2}{\pi}\langle u_iu_j\rangle_{\bu}\partial_i\left(\rho\partial_j\rho\right)=\frac{1}{\pi}\n\cd\left(\rho\n\rho\right)
\eeq
 by using Eqs.~(\ref{isomean1}) and (\ref{isomean2}).

\section{Derivation of the amplitude equations}
\label{Ampderiv}

Here we describe the scheme for the derivation of the
two coupled amplitude equations~(\ref{Aeq})
from the three underlying nonlinear equations~(\ref{fund}).
First of all one assumes small values for the amplitudes  $X$ and $Y$
of the  spatially periodic deviations from the
homogeneous basic state $\rho_0$ and ${\bf t}=0$.
At threshold, i.e. at $\rho_0=\rho_c$ and for $k_c$ as calculated in
Eqs.~(\ref{rhoc}) and ~(\ref{kc}),
these deviations are
either periodic  in $x$- or in $y$-direction
as described by the ansatz in  Eq.~(\ref{ansatz}).

Similar to Sec.~\ref{thresh}, the nonlinear equations (\ref{fund}) may be rewritten
in terms of the deviations
${\bf w} = (\tilde \rho, t_x,t_y)$ from the
basic state ${\bf w}_0=(\rho_0,0,0)$  as follows:
\beq
\label{nlequ}
\p_t {\bf w}
&=&
\mathcal{L}_0 {\bf w}
+{\bf {N}}\left(\rho,\bt\right)\\
{\rm with}\qquad
{\bf {N}}\left(\rho,\bt\right)&=&\begin{pmatrix}\mathcal{N}_{\rho}\left(\rho,\bt\right)\\\mathcal{N}_x\left(\rho,\bt\right)
\\\mathcal{N}_y\left(\rho,\bt\right)\\\end{pmatrix}.
\eeq
The linear operator is defined by Eq.~(\ref{L0entries})
and the  nonlinear operator ${\bf N}$
includes all the nonlinear terms with respect to
$\tilde \rho$ and ${\bf t}$ on the
right hand sides of Eqs.~(\ref{fund}).

Naturally, as the small  expansion parameter
the relative distance from the threshold,
\beq\label{rhoexp2}
\varepsilon = \frac{\rho_0-\rho_c}{\rho_c}~,
\eeq
is chosen. Close to threshold the dynamics of the linear
solution in  Eq.~(\ref{mode}) is
slow and accordingly a slow time scale
\beq
T=\varepsilon t\,
\eeq
is introduced allowing
the time derivatives
in Eqs.~(\ref{fund}) to be  replaced by
\beq\label{timederiv}
\p_t\rightarrow\varepsilon\p_T\enspace.
\eeq
The solution ${\bf w}$ is expanded with respect
to powers of $\varepsilon^{1/2}$
\beq
\label{vexp}
{\bf w}=\varepsilon^{1/2}{\bf w}_1+\varepsilon {\bf w}_2+
\varepsilon^{3/2}{\bf w}_3+ \ldots\enspace,
\eeq
with ${\bf w}_1$ as in Eq.~(\ref{ansatz}),
as are the nonlinearities
\beq
{\bf N}=\varepsilon {\bf N}_1+ \varepsilon^{3/2} {\bf N}_2+ \ldots
\enspace .
\eeq
Sorting the contributions to Eq.~(\ref{nlequ}) with respect to powers
of $\varepsilon$, one ends
up with the following hierarchy of equations
\begin{subequations}
\beq
\varepsilon^{1/2}&:&\mathcal{L}_0 {\bf w}_1=0~,\\
\label{O31}
\varepsilon&:&\mathcal{L}_0 {\bf w}_2=-\mathcal{N}_\rho(\bt_1)e_{\rho}~,\\
\label{O32}
\varepsilon^{3/2}&:&\mathcal{L}_0 {\bf w}_3=\p_T {\bf w}_1-\mathcal{L}_2 {\bf w}_1-\hspace{-2mm}\sum_{i=x,y}\mathcal{N}_i(\rho_2,\bt_1)e_i~,~~
\label{O33}
\eeq
\end{subequations}
that has to be solved successively.
We have introduced $e_{\rho}=(1,0,0)$, $e_x=(0,1,0)$, $e_y=(0,0,1)$  and
\beq
{\cal L}_2=\begin{pmatrix}
\mathcal{L}_{11}^{(2)} & 0 & 0\\
0 & \mathcal{L}_{22}^{(2)} & \mathcal{L}_{23}^{(2)}\\
0 & \mathcal{L}_{32}^{(2)} & \mathcal{L}_{33}^{(2)}\\
\end{pmatrix}
\eeq
with
\beq\label{L2entries}
\frac{1}{\rho_0}\mathcal{L}_{11}^{(2)}&=&\left(\frac{3}{2\pi}-\frac{\al}{24}\right)\Delta
-\frac{19\hspace{1mm}\al}{11520}\Delta^2\,,\nonumber\\
\frac{1}{\rho_0} \mathcal{L}_{22}^{(2)}&=&-\frac{\al}{96}(\Delta+2\p_x^2)
-\frac{\al}{46080}\left(11\Delta^2+64\Delta\p_x^2\right)\,,\nonumber\\
\frac{1}{\rho_0} \mathcal{L}_{23}^{(2)}&=&-\frac{\al}{48}\p_x\p_y-
\frac{\al}{720}\Delta\p_x\p_y\enspace.
\eeq
The remaining two matrix elements $\mathcal{L}_{32}^{(2)}$ and $\mathcal{L}_{33}^{(2)}$ follow
from
 $\mathcal{L}_{23}^{(2)}$ and $\mathcal{L}_{22}^{(2)}$
by permuting
$\p_x$ and $\p_y$.

The equation in $\mathcal{O}(\varepsilon^{1/2})$ is just the linear eigenvalue problem
already discussed as Eq.~(\ref{lineq}) in Sec.~\ref{thresh}
, i.e. it is solved  by
$\tilde \rho_1=0$ and
\beq
t_{1x}=X(T)e^{ik_c x}+c.c.\enspace,\enspace t_{1y}=Y(T)e^{ik_c y}+c.c.
\eeq
describing the orientational fluctuations.

At the next order $\mathcal{O}(\varepsilon)$ , the nonlinearity is only present in
the density equation, while $\mathcal{L}_0$ acting on the $\bt$-subspace is nonsingular,
leading to $\bt_2=0$.
The overall up-down symmetry of Eqs.~(\ref{fund}) with respect to $\bt$ prohibits
hexagonal structures as can be seen in this
order in $\varepsilon$, where for hexagons to be driven
a quadratic contribution in the equations that are linearly unstable, i.e. in the equations for
the orientation fields $t_x$ and $t_y$ would be needed \cite{Ciliberto:90.1}.
Inserting $\bt_1$ in $\mathcal{N}_\rho(\bt_1)$ yields an equation for $\rho_2$,
whose solution is of the following form
\beq\label{rho_2}
\rho_2(X,Y)&=&r_1 X^2e^{2ikx}+r_2 Y^2e^{2iqy}  \nonumber\\
& &\hspace{-8mm}+\hspace{1mm}r_3 XYe^{i(kx+qy)}+ r_4 XY^{*} e^{i(kx-qy)}+c.c.
\eeq
with $r_i=r_i(\alpha,D_r,\gamma)$ for $i=1,..,4$.
Here one can see that the coupling of the orientational field
to the density is crucial for the
physical stability of the system, since the density
$\rho_2$ is responsible for the saturation of the amplitudes in the equation of
the next order $\varepsilon^{3/2}$,  c.f. Eq.~(\ref{O32}), while $\bt_2=0$ and thus $\bt$ can
not limit the amplitudes.

Instead of solving Eq.~(\ref{O33}) at the order $\mathcal{O}(\varepsilon^{3/2})$,
one can use Fred\-holm's al\-ter\-na\-tive, which
states that for Eq. (\ref{O32}) having solutions, there must not exist terms
on its right hand side that lie in the kernel of $\mathcal{L}_0$, i.e. no contributions
proportional to the critical modes $\exp(ik_c x)$, $\exp(ik_c y)$.
Collecting the prefactors of these respective modes, one gets the two equations (\ref{Aeq}),
with analytical but lengthy expressions for $\tau_0,g_1,g_2$ as functions of $\alpha$, $D_r$ and $\gamma$.

\section{Existence and stability of the nonlinear stripe and square state}
\label{squarestab}

The range of existence as well as the range of linear stability for
roll solutions and for the
squares can be easily investigated in terms of the amplitude equations.
Stationary single amplitude solutions
as given in  Eqs.~(\ref{rollsol})
exist beyond threshold, i.e. for $\varepsilon>0$,
only when $g_1>0$ holds.
The amplitudes for a stationary square, Eq.~(\ref{squaresol}),  follow from
Eqs.~(\ref{Aeq}) assuming equal amplitudes
\beq
|X|=|Y|= \sqrt{ \frac{\varepsilon}{g_1+g_2}}\,\,.
\eeq
Obviously, squares exist beyond threshold
only in the parameter range  $g_1+g_2>0$.

\paragraph{Linear stability.}
For small perturbations $\delta X$ and $\delta Y$ of the
stripe and square solutions respectively, one obtains by the ansatz
$X=X_0+\delta X$ and $Y=Y_0+\delta Y$ and linearizing
Eqs.~(\ref{Aeq}) with respect to $\delta X$ and
$\delta Y$  two coupled equations in both
perturbations. Those  may be solved
by the mode ansatz $ (\delta X,\delta Y) \sim (\delta \tilde X, \delta \tilde Y) e^{\sigma t}$
leading to a second
order polynomial in $\sigma$ providing  two eigenvalues. One of them
is always negative while the second is either
\beq
\sigma_r=\varepsilon ~\frac{g_1-g_2}{g_1}
\eeq
for rolls or
\beq
\sigma_s=2\varepsilon~\frac{g_2-g_1}{g_1+g_2}
\eeq
for squares.

Thus stripes or squares are stable
if $\sigma_r$ or $\sigma_s$ is negative, respectively.
Accordingly stripes
are the preferred solution  in the range of the nonlinear coefficients
 $g_2> g_1>0$, while
in the parameter range $|g_2| < g_1$
the square patterns are preferred \cite{Segel:65.1}.
Since the nonlinear coefficients $g_1$ and $g_2$
are functions of the rate of active translational transport $\alpha$
and of the rotational diffusion coefficient
$D_r$, we are able to plot the stability ranges
of the patterns as depicted in Figs.~\ref{phdiag}, \ref{phdiaggam}.

%
%
\bibliographystyle {prsty}
\bibliography{./nlaster}
%

\end{document}